\documentclass{aa}  

\usepackage{graphicx}
\usepackage{txfonts}
\newcommand{\kep}{{\it Kepler }}

\begin{document}

   \title{KIC~4150611: a rare multi-eclipsing quintuple with a hybrid pulsator}

   \author{
	K. G. He{\l}miniak\thanks{Subaru Fellow}\inst{1,2}
	\and
	N. Ukita\inst{3,4}
	\and
	E. Kambe\inst{3}
	\and
	S. K. Koz{\l}owski\inst{1}
	\and
	R. Paw{\l}aszek\inst{1}
	\and
	H. Maehara\inst{3}
	\and
	C.~Baranec\inst{5}
	\and
	M.~Konacki\inst{1}
          }

\institute{
Department of Astrophysics, Nicolaus Copernicus Astronomical Center, ul. Rabia\'{n}ska 8, 87-100 Toru\'{n}, Poland\\ \email{xysiek@ncac.torun.pl}
	\and
Subaru Telescope, National Astronomical Observatory of Japan, 650 North Aohoku Place, 
Hilo, HI 96720, USA
         \and
Okayama Astrophysical Observatory, National Astronomical Observatory of Japan, 3037-5 Honjo, Kamogata, Asakuchi,\\Okayama 719-0232, Japan 
	\and
The Graduate University for Advanced Studies, 2-21-1 Osawa, Mitaka, Tokyo 181-8588, Japan
	\and
Institute for Astronomy, University of Hawai`i at M\={a}noa, Hilo, HI, USA
             }

   \date{Received ..., 2016; ..., 2017}

  \abstract
   {}
{We present the results of our analysis of KIC~4150611 (HD~181469) -- an interesting, 
bright quintuple system that includes a hybrid $\delta$~Sct/$\gamma$~Dor pulsator. 
Four periods of eclipses -- 94.2, 8.65, 1.52 and 1.43~d -- have been observed by the \kep satellite, 
and three point sources (A,B, and C) are seen in high angular resolution images.
}
{From spectroscopic observations made with the HIDES spectrograph attached to the 1.88-m 
telescope of the Okayama Astrophysical Observatory (OAO), 
for the first time we calculated radial velocities (RVs) of the component B -- a pair of G-type stars -- 
and combined them with \kep photometry in order to obtain absolute physical parameters of this 
pair. We also managed to directly measure RVs of the pulsator, also for the first time. 
Additionally, we modelled the light curves of the 1.52 and 1.43-day pairs, and measured their 
eclipse timing variations (ETVs). We also performed relative astrometry and photometry of 
three sources seen on the images taken with the NIRC2 camera of the Keck~II telescope.
Finally, we compared our results with theoretical isochrones.
}
{The brightest component Aa is the hybrid pulsator, transited 
every 94.2 days by a pair of K/M-type stars (Ab1+Ab2), which themselves form a 1.52-day 
eclipsing binary. The components Ba and Bb are late G-type stars, forming 
another eclipsing pair with a 8.65 day period. Their masses and radii are
$M_{Ba}=0.894\pm0.010$~M$_\odot$, $R_{Ba}=0.802\pm0.044$~R$_\odot$ for the primary, and 
$M_{Bb}=0.888\pm0.010$~M$_\odot$, $R_{Bb}=0.856\pm0.038$~R$_\odot$ for the secondary.
The remaining period of 1.43 days is possibly related to a faint third star C, which
itself is most likely a background object.
The system's properties are well-represented by a 35 Myr isochrone, basing on which the masses of 
the pulsator and the 1.52-day pair are $M_{Aa}=1.64(6)$~M$_\odot$,
and $M_{Ab,tot}=0.90(13)$~M$_\odot$, respectively. There are also hints of additional
bodies in the system.
   }
   {}


   \maketitle
%

\section{Introduction}
Stellar astrophysics experiences a renaissance thanks to new, highly-stabilized 
spectrographs and very high precision photometry from space-borne observatories 
like CoRoT, {\it Kepler}, or MOST. The fields that, apart from extrasolar planets,
benefited the most are probably asteroseismology (e.g. numerous discoveries of
solar-type oscillations, or hybrid $\delta$~Sct/$\gamma$~Dor pulsators), and 
eclipsing binaries (e.g. high-precision light curves of thousands of objects, 
precise eclipse timing, multi-eclipsing systems, etc.). The
former allows us to look into the stellar interiors and study the structure of
a star, while the latter brings directly measured, absolute physical parameters of the
studied objects.  Therefore, targets that combine both are extremely important for the
modern astrophysical research.

Another interesting field, which is still relatively poorly studied, are multiple 
systems, especially those with the order of 5 or more. This is mainly due to 
low number of known systems, which decreases rapidly with the number of components. 
The current version of the Multiple Star Catalog 
\citep[MSC;][]{tok97}\footnote{http://www.ctio.noao.edu/$\sim$atokovin/stars}
lists $>$1000 triples, and 220 quadruples, but only 35 quintuples,
13 sextuples, and two septuples (the largest multiplicity order we know). There are
many open questions regarding the formation, dynamical interactions, or even the
abundance of such systems in the Galaxy.

In this work we present our results of a study of an object observed by the \kep
satellite, that combines all the aforementioned properties -- pulsations, eclipses, 
and high-order multiplicity. It is, therefore, one of the most interesting targets
in the field of the original \kep mission.
Since its pulsations have been studied by several authors, here we focus on the 
motion of the components (astrometry, radial velocities, eclipse timing variations) 
and the absolute parameters, in order to infer the evolutionary status and multiplicity 
of the system.

\subsection{The target}

\begin{figure}
\centering
\includegraphics[width=\columnwidth]{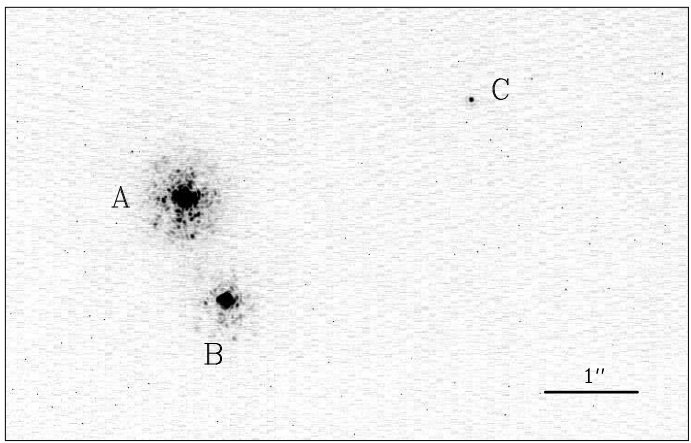}
\caption{Portion of a raw image of KIC~4150611, taken on
2014-07-07 with the NIRC2 camera in the Hcont filter. 
Three point sources are clearly visible. Those marked by A and B
are components of the well-known visual binary
ADS~12310~AB. North is up, and East is left.}\label{fig_ao}
\end{figure}

\begin{figure}
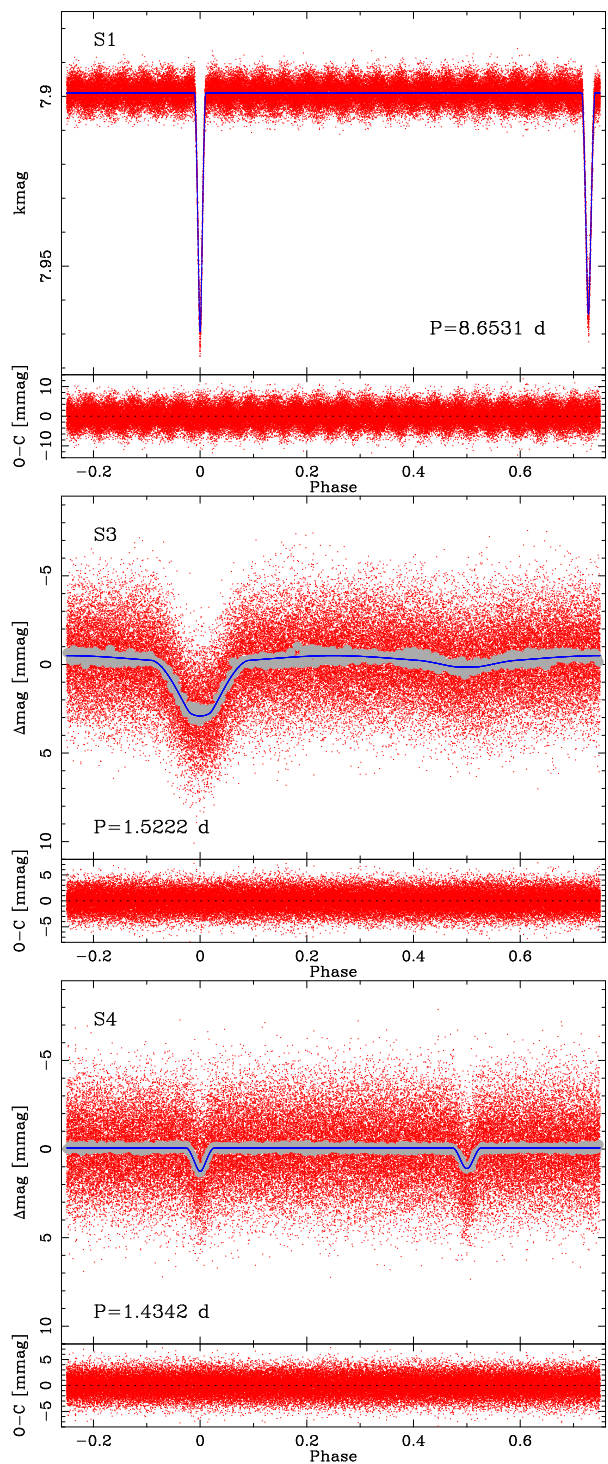

\centering
\includegraphics[width=0.87\columnwidth]{K0415_S1.eps}
\includegraphics[width=0.87\columnwidth]{K0415_S3.eps}
\includegraphics[width=0.87\columnwidth]{K0415_S4.eps}
\caption{The S1 (top), S3 (middle), and S4 (bottom) light curves, representing 
three out of four periods of eclipses observed in the KIC~4150611 system.
Red dots are the observations (S1) or residuals of previous fits cleaned from 
pulsations (S3 and S4). Grey symbols are bins of 200 points in phase domain. 
Blue lines are the JKTEBOP models. Please note the change of the spread of 
residuals of each fit (lower panels). The respective parameters are listed in 
Table~\ref{tab_0415_kk}.
}\label{fig_mod_0415}
\end{figure}

\begin{table}[h!]
\centering
\caption{Basic literature information on KIC~4150611}\label{tab_prop}
\begin{tabular}{lcc}
\hline \hline
Parameter & Value & Reference \\
\hline
$\alpha_\mathrm{ICRS}$ (J2000)&  $19^\mathrm{h} 18^\mathrm{m} 58\fs209$ &	1 \\
$\delta_\mathrm{ICRS}$ (J2000)& $+39^\circ 16' 01\farcs687$ &	1 \\
$\mu_\alpha$ (mas/yr)		& $-6.949(35)$	& 1 \\
$\mu_\delta$ (mas/yr)		& $-5.962(38)$	& 1 \\
$\varpi$ (mas)			& 7.73(46)	& 1 \\
Sp. Type			& F1 V mA9	& 2 \\
\multicolumn{2}l{\bf Observed magnitudes:}& \\
$B$ & 8.29 & 3 \\
$V$ & 8.00 & 3 \\
\kep & 7.899 & 4 \\
$J$ & 7.185(39) & 5 \\
$H$ & 7.029(38) & 5 \\
$K$ & 6.945(24) & 5 \\
\hline
\end{tabular}
\\References: (1) {\it Gaia} DR1 \citep{gai16};\\
(2)~\citet{nie15};
(3)~{\it Simbad} \citep{wen00}; 
(4)~\kep Input Catalog;
(5)~2MASS \citep{skr06}
\end{table}

Having the observed magnitudes of 7.899 in the \kep band and 8.00 in
Johnson's $V$, the \object{KIC 4150611} (a.k.a. KOI~3156, HD~181469, HIP~94924, ADS~12310~AB,
WDS~J19190+3916AB) is one of the brightest objects listed in the \kep Eclipsing Binaries 
Catalog \citep[KEBC;][]{prs11,sla11,kir16}\footnote{http://keplerebs.villanova.edu/}.
It is usually classified as having a late-A spectral type. Its basic properties are 
summarised in Table \ref{tab_prop}.

According to the Washington Double Star catalogue, \citep[WDS;][]{mas01}, it is a 
visual binary, discovered by F. Struve (no particular reference is given). 
The WDS currently holds 20 position measurements, with first one dated back to
1831, and the last one taken in 2015. During this time, the separation between the 
components A and B changed from $\sim$1$\farcs$8 to $\sim$1$\farcs$12, 
and the position angle from $\sim$206$^\circ$ to $\sim$203$^\circ$.

The eclipses were first noted in the KEBC, where three periods are currently given:
1.5222786(19), 8.6530923(36) and 94.1982(7)~d. The second value was first found in
\citet{prs11}, and is incorrectly attributed to the brighter component in the WDS.
Additional eclipses were reported later by \citet{sla11}, and also by \citet{oro15}.
Multiple eclipse periods are first given by \citet{row15}, who claim four values:
0.7611224(29), 1.4342003(58), 8.6530988(66), and 94.22454(40)~d, which can be found in
the \kep Object of Interest (KOI) data base. The first one is 
obviously half of the 1.522~d value from KEBC, while the second one (1.434~d) 
does not appear there.

KIC~4150611 was first found to be a spectroscopic binary by \citet{mol11}, but they acquired 
only three spectra, and did not relate the spectral lines to any particular component, 
nor gave the orbital solution. They were also the first to mention this object in relation 
to pulsations. Shortly after it was identified as containing a hybrid $\delta$~Sct/$\gamma$~Dor 
pulsator by \citet{uyt11}. \citet{shi12} studied the $\delta$~Sct pulsations, and used
their new Fourier-transform-based formalism to obtain a radial velocity (RV) curve
of the pulsating component, with the orbital period of $94.09\pm0.11$~d. Later, \citet{bal14}
reconstructed a somewhat different RV curve, with $P=94.3\pm0.1$~d, using a different approach. 
Finally, \citet{nie15} performed spectral analysis and obtained atmospheric parameters of the
pulsator, showing that it is a rapidly-rotating star ($v\sin(i)=128\pm5$~km/s)
of spectral type F1~V~mA9 ($T_{eff}=7400\pm100$~K).


\section{Data and methodology}

\subsection{Keck II/NIRC2 adaptive optics imaging}

\begin{table*}
\centering
\caption{The log and results of Keck~II/NIRC2 observations. We show the date, filters 
(combination of two wheels), 
total number of exposures ($N_{exp}$), integration time per exposure ($t_{int}$), and 
number of co-adds per exposure. In the lower part of the table we show angular separations
($\rho$) and position angles ($\theta$) of stars B and C relatively to A, and
their magnitude differences.}\label{tab_ao_obslog}
\begin{tabular}{lllllll}
\hline \hline 
MJD & Date & Filter & $N_{exp}$ & $t_{int}$ & Co-adds & PI \\ 
 & & combination & & [s] & & \\
\hline 
56458.61442	& 2013-06-15	& PK50\_1.5+Br$\gamma$	& 15 &  1.0 &  5 & Marcy	\\
56845.42700	& 2014-07-07	& PK50\_1.5+Kcont	&  6 & 12.5 & 25 & Knutson	\\
56845.43081	& 		& PK50\_1.5+Jcont	&  6 & 12.5 & 50 &		\\
56845.43408	& 		& PK50\_1.5+Hcont	&  6 & 12.5 & 50 &		\\
57643.28633$^a$	& 2016-09-11	& PK50\_1.5+Kcont	&  5 & 12.5 & 25 & Baranec	\\
57677.24852	& 2016-10-15	& PK50\_1.5+Kcont	& 8$^b$& 30/60& 30/60&Baranec	\\
57677.25537	&		& K+clear 		& 3$^c$& 1.81/18.1& 10/100 &	\\
\hline \hline
MJD & $\rho_{AB}$ & $\theta_{AB}$ & $\rho_{AC}$ & $\theta_{AC}$ & $\Delta mag_{AB}$ & $\Delta mag_{AC}$ \\
& [mas] & [$^\circ$] & [mas] & [$^\circ$] & [mag] & [mag] \\ 
\hline
56458.61442	& 1151.51(39) & 203.279(16) & 3279.7(3.8) & 287.794(71) & 1.087(5) & 5.00(13) \\ 
56845.42700	& 1146.17(11) & 203.503(12) & 3266.48(33) & 288.002(13) & 1.104(5) & 5.139(47) \\ 
56845.43081	& 1146.24(9)  & 203.509(13) & 3267.28(67) & 287.987(15) & 1.368(7) & 5.266(55) \\ 
56845.43408	& 1146.42(8)  & 203.515(11) & 3267.47(69) & 287.990(15) & 1.133(7) & 5.168(22) \\ 
57643.28633$^a$	& \it 1131.76(20) & \it 203.053(36) & \it 3226(12)&\it 287.862(47)& \it 1.099(7) & \it 5.3(2) \\
57677.24852	& 1134.30(22) & 203.490(22) & 3242.38(52) & 288.284(20) & 1.125(5) & 5.180(30) \\ 
57677.25537	& 1134.67(39) & 203.468(22) & 3243.17(87) & 288.290(26) & 1.052(2) & 5.053(34) \\
\hline 
\end{tabular}
\\$^a$ Observations with bad AO correction. Results were not used in the further analysis.
\\$^b$ The first two frames are composed of 30 co-adds (1~s each), while
the following six are composed of 60 co-adds (also 1~s).
\\$^c$ The first frame is composed of 10 co-adds (0.181~s each), while 
the following two are composed of 100 co-adds (also 0.181~s).
\end{table*}

In the Keck Observatory Archive (KOA)\footnote{https://koa.ipac.caltech.edu/} we
have found high-angular-resolution images of KIC~4150611, obtained with the
NIRC2 camera, which is fed by the natural guide star adaptive optics 
\citep[NGS~AO;][]{wiz00} on the Keck~II telescope. These observations were taken 
on 2013 June 15 (filter: Br$\gamma$; program ID:~U078N2; PI:~Marcy) and 2014 
July 07 (filters: Jcont, Hcont, Kcont; program ID:~C191N2; PI:~Knutson).
We have supplemented these data with our own observations, performed on 2016 
September 11 in the Kcont filter, and 2016 October 15 in filters Kcont and K.
Unfortunately, on our first night the conditions were bad, which caused a poor AO correction. 
All observations were done in the ``narrow'' mode, which gives the field of view of 
$\sim$10$'\times10'$. As usual for infra-red observations, dithering was performed
in order to remove the flux of the sky in further processing.
Table~\ref{tab_ao_obslog} shows the observing log for AO imaging.

Except for the Br$\gamma$ data, from the KOA we downloaded raw frames, and we
reduced and analysed them by ourselves with a combination of IRAF and Python-based routines. 
Under IRAF we applied corrections for bad pixel, flat field, and sky flux.
Because there are no master flats for the -cont filters available from the NIRC2 website, 
we took our own calibrations for Kcont and Hcont, while for the Jcont band we had to look 
in the KOA for appropriate calibrations. In case of Br$\gamma$ frames, we used
the calibrated ones, available in KOA directly. All data were then corrected for
distortion. This was done with a freely-available, dedicated Python 
script\footnote{https://github.com/jluastro/nirc2\_distortion/wiki} 
which was prepared to follow the prescription described in \citet{yel10}
and \citet{ser16}, and which utilizes distortion maps that are available on line. 
Note, that the maps for observations taken after 2015 April differ from those 
from before that date. Therefore we used different
maps for the KOA (from 2013 and 2014) and our own data (2016).

The images show three point sources: two bright ones, the A and B components
that form the visual binary ADS~12310~AB, and a faint one (C), located 
approximately 3 arcsec NW from the bright pair (Figure~\ref{fig_ao}). This
star was out of the field of view of some 2013 data, due to dithering.
We rejected such frames from our analysis.

For all three sources we performed relative astro- and photometric measurements. 
The positions of stars on the chip ($x,y$) were measured by fitting a 2D Gaussian
to the core of the star's point spread function (PSF). They were then translated to 
distances along the chip's axes in angular scale ($\Delta \hat{x}$,$\Delta \hat{y}$), 
and later, to relative angular separations and position angles ($\rho$, $\theta$). 
For this we used the pixel scale and rotation angle values given on the same website 
as the distortion script, coming from \citet{yel10} and \citet{ser16}. They are:
$9.952\pm0.002$~mas/pix and $0\fdg252\pm0\fdg009$ for data taken before, and
$9.971\pm0.004$~mas/pix and $0\fdg262\pm0\fdg020$ for data taken after 2015 April. 
Please note that we did not de-rotate the images, only applied the correction to
the measurements of $\theta$.

Finally, we applied corrections for the atmospheric refraction, using the method described 
in \citet{hel09a}, assuming monochromatic refraction at the effective wavelength of a 
given filter, and calculating the refractive index $n$ with the method of \citet{mat04,mat07}. 
For the atmospheric conditions (outside temperature, air pressure and humidity) we
either used the values listed in the headers of files downloaded from KOA, or
(for our own observations) we checked them in the KOA Ancillary Weather Data 
service\footnote{https://koa.ipac.caltech.edu/UserGuide/ancillary.html}. These corrections
were, however, smaller than the measurement errors.

The instrumental fluxes were also measured by fitting a 2D Gaussian, but the 
photometry was done on frames not corrected for distortion. To obtain the magnitude
differences we simply translated the flux ratios using the standard formula
$\Delta mag = -2.5\log(F_2/F_1)$.

All individual measurements were averaged, and $rms/\sqrt{N_{exp}}$ was taken as the 
uncertainty. The results are shown in Table~\ref{tab_ao_obslog}. Very large errors 
for 2016 September measurements are caused by poor AO correction. That night, 
the star C was extremely difficult to detect, therefore its measurements are highly 
uncertain, and we did not use them in the further analysis. Also, due to short integration 
times and narrow bandwidth of the filter, this star was barely detectable on the Br$\gamma$
images. These measurements also have relatively large error bars.

The main purpose of our own AO observations was: (1) to determine if the star C is 
gravitationally bound to the AB pair, and (2) if it shows eclipses. These are discussed in
further sections of this paper.

\subsection{\kep photometry and light curve analysis}\label{sec_kep_phot}

\begin{table*}
\centering
\caption{Results of the JKTEBOP fit to the S1, S3, and S4 curves, and absolute
magnitudes in the \kep band.}\label{tab_0415_kk}
\begin{tabular}{lccc}
\hline \hline
Curve & S1 & S3 & S4 \\
\hline
$P_{ecl}$ (d) 		& 8.6530941(16)	& 1.43420486(12)& 1.5222468(25) \\
$T_0$ (JD-2454900)	& 61.00508(17)	& 60.76059(7) 	& 60.8799(14) \\
$e$ 			& 0.374(7) 	& 0.0(fix) 	& 0.0(fix) \\
$\omega$ ($^\circ$)	& 13.0(2.6) 	& --- 		& --- \\
$r_1$ 			& 0.0373(21) 	& 0.328(18) 	& 0.093(16) \\
$r_2$			& 0.0398(17) 	& 0.212(38) 	& 0.071(18) \\
$i$ ($^\circ$) 	& 89.28(14) 	& 88.5$^{+0.5}_{-6.4}$ & 88.8$^{+1.2}_{-2.6}$\\
$J$ 			& 0.943(42) 	& 0.111(22)	& 0.934(16) \\
$L_2/L_1$ 		& 1.07(17) 	& 0.046(26) 	& 0.5$^{+0.7}_{-0.3}$\\
$L_3/L_{tot}$ 		& 0.8579(45) 	& 0.9936(9) 	& 0.997(2)\\
$L_1/L_{tot}$ 		& 0.06865(61)	& 0.00612(87)	& 0.0020(14)	\\
$L_2/L_{tot}$ 		& 0.07245(60)	& 0.00028(16)	& 0.0010$_{-0.0008}^{+0.0011}$	\\
$Kmag_1$ (mag)$^b$	&  5.25(13)	&  7.87(20)	&  9.09(77)\\
$Kmag_2$ (mag)$^b$	&  5.19(19)	& 11.22(62)	&  9.84$^{-1.20}_{+0.88}$ \\
$rms_{LC}$ (mmag) 	& 3.13 		& 2.30/0.21$^c$	& 1.86/0.09$^c$ \\
\hline
\end{tabular}
\\$^a$ From the periodogram analysis. 
$^b$ From the observed magnitude of the whole system (7.899~mag)
\\ and {\it Gaia} DR1 distance ($129.4\pm7.7$~pc).
$^c$ For the unbinned and binned curve, respectively.
\end{table*}

In this study we make use of the Q0-Q17 \kep mission photometry, publicly 
available from the KEBC website. We used the de-trended relative flux measurements $f_{dtr}$, 
that were later transformed into magnitude difference $\Delta m=-2.5\log(f_{dtr})$,
and finally the KEBC value of $k_{mag}$ was added. We call the resulting light
curve (LC) the {\it stage~0} (S0) curve. Due to the amount of data and limited computational 
resources, only long-cadence data were used in the LC analysis.

Before the light curve (LC) analysis, we filtered out the eclipses of the F1-type star, 
applying the following ephemerides: $2544840.707336 + E\times 94.226$~d, which are 
slightly different than those given in the KEBC, but in agreement with \citet{bal14}
or \citet{row15}. With this procedure we also removed several eclipses that 
occurred in the component B, and which coincided with the 94-day period. 
We called this resulting light curve the {\it stage~1} (S1) curve. The main
variability in this curve are the eclipses with 8.65-day period.

For all the LC fits we used version 28 (v28) of the code JKTEBOP 
\citep{sou04a,sou04b}, which is based on the EBOP program \citep{pop81}. 
We fitted for the period $P$, primary (deeper) eclipse mid-time $T_0$, 
eccentricity $e$, periastron longitude $\omega$, inclination $i$, 
ratio of fluxes $L_2/L_1$, ratio of central surface brightnesses $J$, 
sum of the fractional radii $r_1+r_2$ (in units of major semi-axis $a$), 
their ratio $k$, and fractional amount of the third light $L_3/L_{tot}$.
With several other sources of brightness variation (pulsations, additional 
eclipses), the S1 curve can be treated as affected by a correlated (red) noise.
Therefore, for reliable error estimation, we applied the residual-shifts (RS) 
method. The best fit was done on the complete Q0-Q17 long-cadence light curve, 
but, due to amount of data, for the error estimation we worked on single-quarter LCs.
Our approach is common for all \kep targets from our program and is described in 
details in \citet{hel16}.

We first made the fit for the 8.65-day pair to the S1 curve, and run the RS to 
calculate the uncertainties for this system. We then took the residuals and run 
a Lomb-Scargle (LS) periodogram\footnote{Lomb-Scargle periodograms for this work were 
created with the on-line NASA Exoplanet Archive Periodogram Service:\\ 
http://exoplanetarchive.ipac.caltech.edu/cgi-bin/Pgram/nph-pgram.}  
in order to identify additional periods. 

We found a number of peaks related to pulsations, as well as a period of eclipses 
of 1.434205~d (see Sect.~\ref{sec_per_1.43}). We run the JKTEBOP once again with this 
period fixed, only to correct for these eclipses and improve the identification of 
frequencies of pulsation. The residuals of this intermediate JKTEBOP fit we call the 
{\it stage~2} (S2) curve. We made another periodogram run on it, and we identified 
the most prominent pulsation periods (Sect.~\ref{sec_hyb}). We then removed the 
pulsations from the {\it residuals of the S1 curve}, and obtained the {\it stage~3} 
(S3) curve, in which the main variability are the eclipses of the 1.43-day period. 
The final JKTEBOP fit for this period, the second one with proper error calculations, 
was done on the S3 curve. The residuals of this fit to S3 we call the {\it stage 4} 
(S4) curve, in which we can see eclipses with the 1.52-day period. The last JKTEBOP 
run was done on the S4 curve.

Because the pulsations were mainly removed, the phase-folded S3 and S4 curves show
a significantly smaller noise than the residuals of S1. We treated S3 and S4 as
showing no correlated noise, and used a bootstrap approach to calculate
the uncertainties. 

We present the results of JKTEBOP fits to curves S1, S3, and S4 in 
Table~\ref{tab_0415_kk} and Figure~\ref{fig_mod_0415}. Apart from the parameters 
given directly by JKTEBOP, we also calculated the fractional fluxes of each 
component and used them together with the apparent magnitude of the system 
to calculate individual apparent magnitudes in the \kep band ($kmag$). 
Then, using the known distance, and assuming it is the same for all three pairs, 
we estimated the absolute ones ($Kmag$). The distance modulus is 
$Kmag-kmag=-5.56\pm0.13$ mag. Zero extinction is assumed. The sum of
the individual fractional fluxes also allows us to estimate the contribution
from the remaining component A to be 84.95 per cent and its absolute magnitude
$Kmag_A=2.52(13)$~mag (assuming that the total light in the \kep LC comes
only from seven stars: the pulsator and components of three shorter-period EBs).

\subsection{HIDES spectroscopy and radial velocities}

\begin{figure}
\centering
\includegraphics[width=\columnwidth]{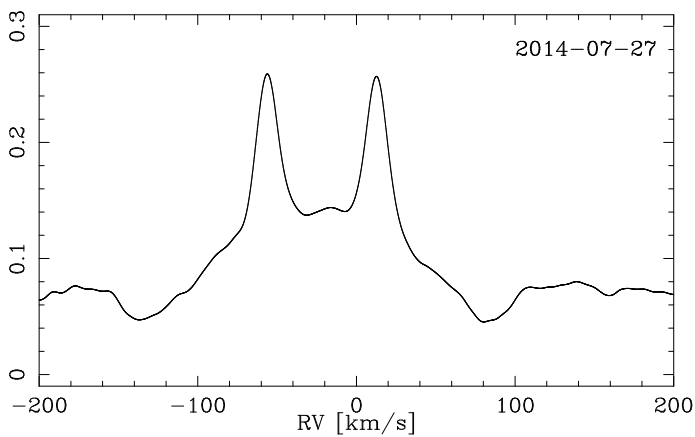}
\caption{Cross-correlation function of KIC~4150611, from the
spectrum taken on July 27, 2014. Two sharp peaks correspond to
the G-type stars forming the 8.65-day binary (components Ba+Bb), 
the broad one is due to the F1-type star (component Aa). Velocities 
not shifted to the Solar system barycentre.}\label{fig_ccf_0415}
\end{figure}

We observed the target as part of our program of spectroscopic monitoring
of bright \kep eclipsing binaries \citep{hel15,hel16,hel17}. We used the 1.88-m
telescope of the Okayama Astrophysical Observatory (OAO-1.88), with the
HIgh-Dispersion Echelle Spectrograph \citep[HIDES;][]{izu99}, fed through a
2\farcs7 diameter circular fibre \citep{kam13}. Light from both visually separated 
components was collected. The instrument set-up, observational strategy,
and spectroscopic data reduction scheme are described in details in \citet{hel16}.

Between July 2014 and October 2016, we obtained 17 high resolution
($R\sim50000$) spectra. Despite four eclipsing periods were detected in \kep data, 
velocities corresponding to only two of them could be measured -- of a pair of G-type 
stars on a 8.65-day orbit, and of the F1-type pulsator on the 94.2-day orbit.
The peculiar cross-correlation function (CCF) is shown in Figure \ref{fig_ccf_0415}. 
It is composed of two sharp peaks, that belong to the components of the G-type binary, 
and a very broad one, coming from the fast rotator of spectral type F1. This 
contaminator introduces additional systematic variation to the RV measurements of the 
other two components visible in spectra, and is the probable source of systematic 
uncertainties that dominate the error budget of their orbital fit. 

For calculation of the RVs of the G-type pair we decided to use our own implementation 
of the TODCOR technique \citep{zuc94}, which finds velocities $v_1$ and 
$v_2$ of two stars simultaneously. Individual RV errors were estimated with a 
bootstrap approach \citep{hel12}.

The RVs of the F1-type star (component A) were measured from the position
of the $H_\beta$ line (4861.363~\AA). We chose this line due to the
fact that the Balmer series are the dominant features in the spectra, and
$H_\beta$ lays in a spectral range that is significantly less affected by 
lines from the G-type pair and tellurics than, for example, $H_\alpha$. It is
also in the centre of its echelle order, and the SNR around it is significantly
higher than around lines at shorter wavelengths, like $H_\gamma$ or $H_\delta$. 
We measured its position by fitting a Gaussian to its core under IRAF's task
{\it splot}. As the measurement error, we assume a conservative value of 
0.1~\AA, which corresponds to $\sim$6 km/s at this wavelength. The orbital
fit later showed that these errors were in fact overestimated. 

\subsection{RV orbital fits}

\begin{figure*}
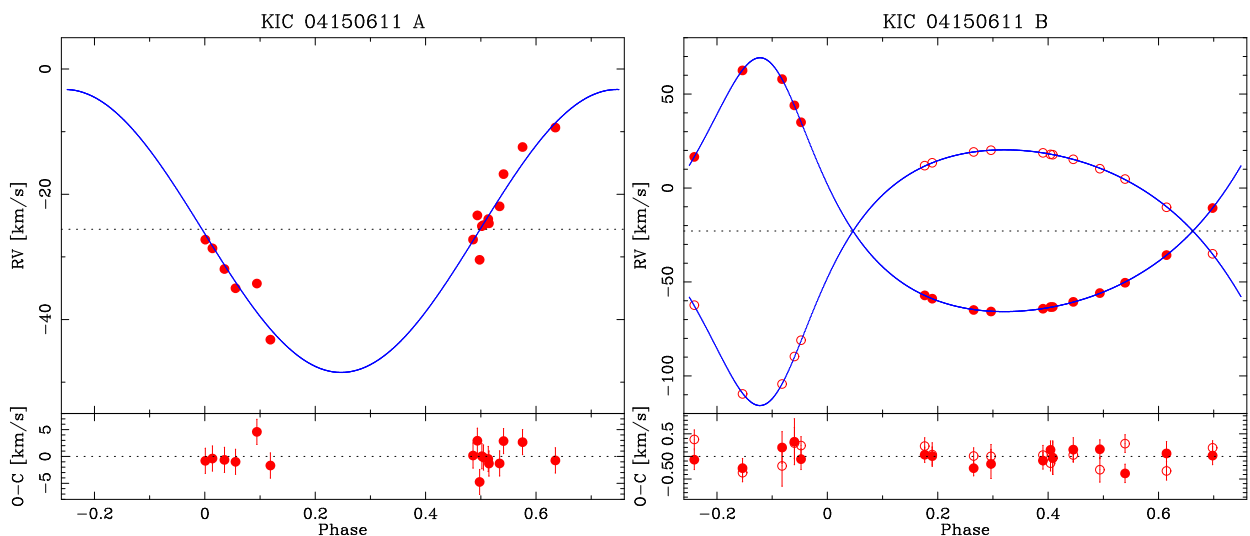

\centering
\includegraphics[width=0.9\columnwidth]{K0415A_orb.eps}
\includegraphics[width=0.9\columnwidth]{K0415B_orb.eps}
\caption{Radial velocity curves of the components A (left) and B (right) of KIC~4150611, 
phase folded with their orbital periods. The best-fitting models are plotted with 
blue lines. Filled circles on the right panel refer to the primary Ba, 
and open ones to the secondary Bb. Phases 0 are set for the eclipse mid-times
(the deeper one in case of B).
}\label{fig_rv_0415}
\end{figure*}
 
\begin{table}
\centering
\caption{Results of the V2FIT fits to the RV measurements of the
F1-type pulsator (component A), and the G-type pair (component B).}\label{tab_par_sb2}
\begin{tabular}{lcc}
\hline \hline
Parameter	& Component A & Component B \\
\hline
Type of fit		& SB1		& SB2	\\
$P$ (d)			& 94.226(fix)	& 8.6530941(fix)	\\
$T_p$ (JD-2454900)	& 105.5(1.0)	& 60.0781(28)	\\
$K_1$ (km/s)		& 22.6(2.2)	& 67.55(19)	\\
$K_2$ (km/s)		&    ---	& 68.04(21)	\\
$e$ 			& 0.0(fix)	& 0.374(fix)	\\
$\omega$ ($^\circ$)	&    ---	& 12.91(17)	\\
$\gamma$ (km/s)	& -25.86(62)	& -22.882(38)	\\
$a_1\sin(i)$ (R$_\odot$)& 42.1(4.0)	& 10.714(32)	\\
$a_2\sin(i)$ (R$_\odot$)& ---		& 10.792(34)	\\
$f(m)$ (M$_\odot$)	& 0.113(33)	&  --- \\
$M_1\sin^3(i)$ (M$_\odot$)	&  --- 	& 0.894(10)	\\
$M_2\sin^3(i)$ (M$_\odot$)	&  --- 	& 0.887(10)	\\
$rms_1$ (km/s)		& 2.19		& 0.18	\\
$rms_1$ (km/s)		& ---		& 0.23	\\
\hline
\end{tabular}
\end{table}

The RV orbits were found with the code V2FIT \citep{kon10}, which fits a
single or double-keplerian orbit with a Levenberg-Marquardt algorithm.
We performed two orbital fits to the measured RVs. First for the 8.65-day
pair (component B), treated as a double-lined spectroscopic binary (SB2), 
and second for the F1-type pulsator on a 94.2-day orbit (component A)
treated as a single-lined spectroscopic binary (SB1).
In the SB2 fit we fixed the period and eccentricity to the values found
with JKTEBOP. We fitted for velocity amplitudes ($K_1,K_2$), longitude
of pericentre ($\omega$), systemic velocity ($\gamma$), and moment of
the pericentre passage ($T_p$). With V2FIT it is possible also to fit
for the difference between systemic velocities of two components 
($\gamma_2-\gamma_1$), but we initially found it undistinguishable from
zero.

In the SB1 fit, we set the period to the value that we used to filter out the 
eclipses, 94.226~d, and also kept it fixed. When set free, the result was nearly
the same, but did not reproduce the moments of eclipses very well. Other 
parameters were essentially the same in both cases. We also kept $e$ fixed, 
as it was initially found indifferent from zero; i.e. we found $e<0.17$. 
This agrees with \citet{shi12}, who found it smaller than 0.12, and with \citet{bal14}, 
who gives the value of 0.043, which is undetectable in our data. Other 
parameters were essentially the same, no matter if $e$ was fixed, or 
fitted for. In the final fit, we were only looking for $K$, $\gamma$ and
$T_p$, which for circular orbits is defined as the moment of the first
quadrature (highest value of RV). The value of $\gamma$ in this case should 
be taken with some caution, as it depends on the exact reference wavelength 
that was used in RV measurement. This, however, has no impact on the velocity 
amplitude.

Results of both fits are summarised in Table~\ref{tab_par_sb2}. Observed and
modelled RV curves are shown in Figure~\ref{fig_rv_0415}. Uncertainties
were estimated with a bootstrap procedure, which properly accounts for
possible systematics. The individual RV measurement errors were re-scaled,
so the final reduced $\chi^2$ is close to 1.

\subsection{Eclipse timing variations}\label{sec_etv}
Two of our LCs -- the S3 and S4 curves -- were also checked for the eclipse 
timing variations (ETVs). The 8.65-day eccentric binary (our S1 curve) has been 
analysed by \citet{bor16}, who did not detect any significant signal.

We used the radio-pulsar-style approach, presented in \citet{koz11}. In this 
method, a template LC is created by fitting a trigonometric (harmonic) series 
to a complete set of photometric data. Then, the whole set of photometric data 
is divided to a number of subsets. Their number is arbitrary, but for this study
we set it to 200. For each subset, the phase/time shift is found by fitting 
the template curve with a least-squares method. This approach is well suited 
for large photometric data sets, especially those obtained in a regular cadence, 
like from the \kep satellite.


\section{Results}

\subsection{The hybrid pulsator and its motion}\label{sec_hyb}

\begin{figure*}
\centering
\includegraphics[width=0.99\textwidth]{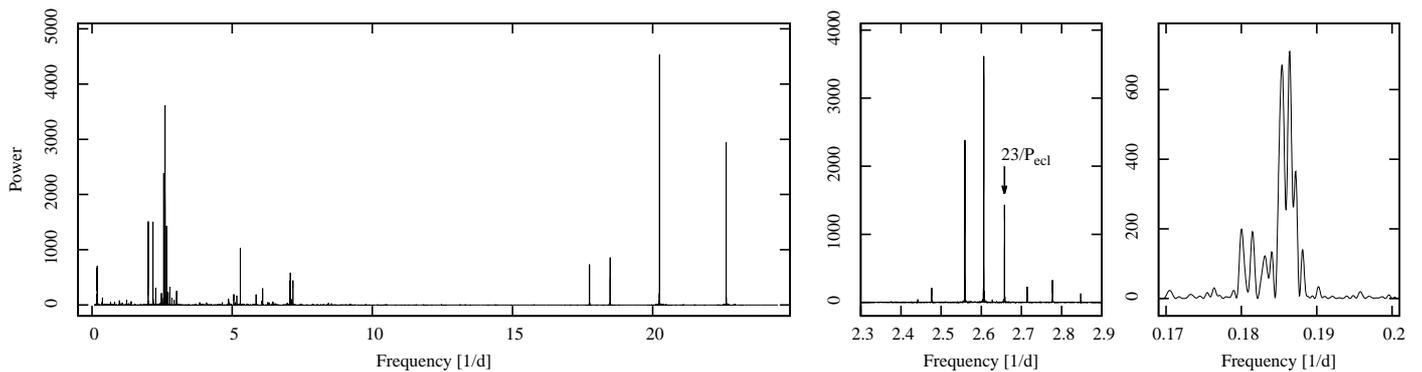}
\caption{{\it Left}: An LS periodogram of the residuals of the LC fit of KIC~4150611, 
with the second period of eclipses (1.43421~d) removed. Peaks representing 
$\delta$~Sct and $\gamma$~Dor pulsations of the component Aa are seen at 
frequencies $>$16 and $\sim$2.6~d$^{-1}$, respectively. 
{\it Middle}: Zoom on the frequencies around the highest peak of the 
$\gamma$~Dor pulsations. The arrow marks the 23-rd harmonic of the 8.65-d period 
(23/8.653094~d$^{-1}$). {\it Right}: Zoom on the frequencies corresponding to 
the rotation period. Their structure suggest differential rotation and/or 
that they come from two stars.
}\label{fig_per0415}
\end{figure*}

In the residuals of the first fit we see several periodicities, which are
even clearer in the S2 curve. The most prominent are the $\delta$~Scuti 
type pulsations of the F1-type star (highest peak at 20.243~d$^{-1}$), also 
reported by \citet{shi12} and \citet{bal14}, and the $\gamma$~Dor type 
pulsations (highest peak at 2.6064~d$^{-1}$). The $\delta$~Sct and $\gamma$~Dor 
pulsations have similar amplitudes, and the star meets the criteria of being a 
hybrid pulsator \citep{bra15}. The corresponding periodogram is presented 
in Figure~\ref{fig_per0415}. A zoom on the $\gamma$~Dor area reveals
a peak that coincides with the 23-rd harmonic of the 8.65-days orbital period.
This mode is clearly seen in the phase-folded S1 curve (Fig.~\ref{fig_mod_0415}; 
top), but it is rather a coincidence than a pulsation mode induced in one of the
G-type stars during close periastron passages, as it takes place in ``heartbeat'' 
(HB) stars \citep{bec14}. No such oscillations have been observed in main sequence
stars of this type, and in the HB stars they usually have a declining amplitude.
The exact frequency $f_{23}=2.6578921$~1/d correspondst to the period of 0.376238~d.
Multiplied by 23, it gives 8.653474~d, which is close to but significantly different 
than $P_B=8.6530941(16)$~d. The S1 curve phase folded with the period of 8.653474~d 
looks much worse than with the best-fitting period.

Another small but clearly seen group of peaks can be found at frequencies 
around 0.185~d$^{-1}$, shown on a zoom in Fig.~\ref{fig_per0415}. They most
likely come from rotation of one of the components (probably the pair B), 
and their complicated structure reveals that the rotation is differential,
and/or that they may be coming from two stars.

Both \citet{shi12} and \citet{bal14} used the pulsations to detect the orbital 
motion of the F1-type component with the 94.2-day period. They used different 
approaches, which resulted in different values of $a\sin(i)$ and the predicted
velocity amplitude $K$. \citet{shi12} used the split of periodogram peaks of 
pulsation in the frequency domain, caused by the orbital motion, but 
\citet{bal14} argues that such side-lobes of the main peaks may be obscured or 
distorted by other, independent pulsation modes of similar frequencies. He also 
notes that amplitude variations may also generate side lobes in the frequency 
domain, and decided to work with all frequencies, and look for time delay. 

Closer inspection of both results is actually confusing. \citet{shi12} clearly
miscalculated (or made a typo in) the values of $a_1\sin(i)$ from their Table~4, 
giving them close to 1.2~AU, while the value expected from their RV amplitudes 
($K\simeq23.5$~km/s) would be rather 0.2~AU. Similar value can be deduced from 
Figure~15 of \citet{bal14}, but in his Table~3, he gives $a_1\sin(i)=0.140$,
and a (properly) corresponding $K=16.2$~km/s. This solution is in agreement
with his measurements of pulsation time delay, presented later in Figure~16.

Thanks to our direct RV measurements, and the orbital solution 
(Table~\ref{tab_par_sb2}), we can clarify the situation. Our value of 
$K_1=22.6\pm2.2$~km/s agrees better with the one of \citet{shi12}, and the 
corresponding $a_1\sin(i)$ is almost exactly 0.2~AU ($0.196\pm0.019$). The value 
that can be deduced from Balona's Figure~15 is therefore correct. 

\subsection{Astrometry of the AB pair}\label{sec_ao_AB}

\begin{figure}
\centering
\includegraphics[height=7cm]{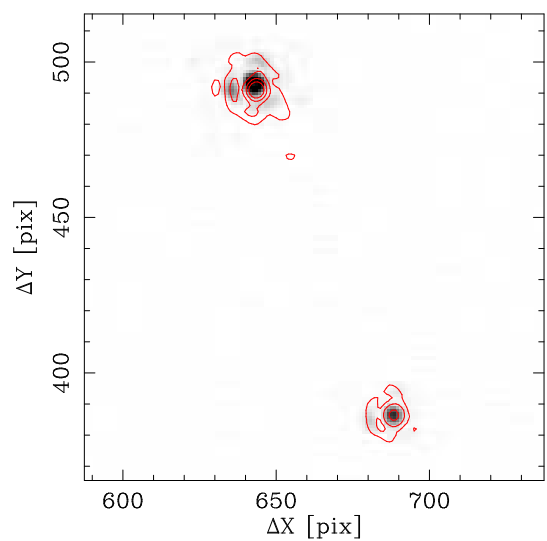}
\caption{Portion of distortion-corrected images in Br$\gamma$ from 2013-06-15
(halftone) and Kcont from 2016-10-15 (red contours). Images are shifted to match
positions of the component B. The centres of the component's A PSF are misplaced
by $\sim$2~pix (nearly 20~mas, or 2.6~AU at the distance to the system).
}\label{ao_fits}
\end{figure}

\begin{figure*}
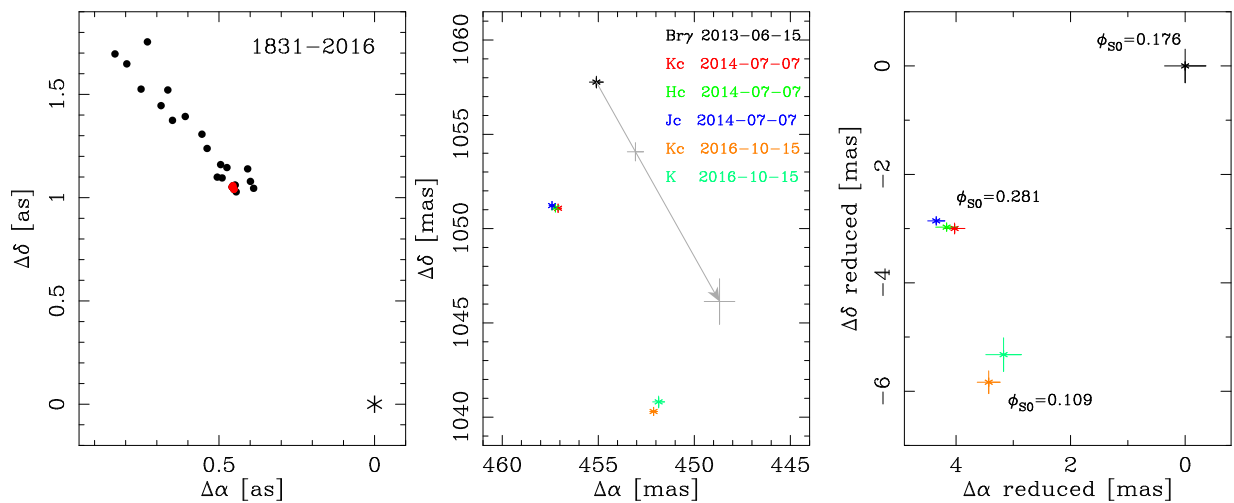

\centering
\includegraphics[height=6.5cm]{AOastro_AB_lit.eps}
\includegraphics[height=6.5cm]{AOastro_AB_nar.eps}
\caption{Astrometry of star A relatively to star B. 
{\it Left:} All data available from the WDS (black) and results from this work (red), 
on $\Delta\alpha/\Delta\delta$ plane, reconstructed from $\rho$ and $\theta$.
The star B is set in (0,0) and marked with an asterisk.
One can see a gradual movement with an average rate of 1.9(2) mas/yr and 3.5(3) mas/yr 
in $\alpha$ and $\delta$ respectively.
{\it Middle:} Zoom onto our results, showing Keck~II/NIRC2 observations (both archival and
ours). Each measurement (date, filter) is denoted with a different colour. The grey arrow
and crosses show the vector of the gradual orbital movement since 2013-06-15 and predicted 
relative positions in dates of other observations (with uncertainties).
{\it Right:} Same measurements, but corrected for the gradual motion, and shifted, so the
first Keck point is in (0,0). Labels $\phi_{S0}$ show phases of the 94.2-day orbit, according
to the ephemeris used to clean the S0 curve from eclipses (Sect.~\ref{sec_kep_phot}). 
The data are clearly inconsistent with a $P=94.2$~d, $\hat{a}\simeq1.55$~mas,
$i\simeq90^\circ$ orbit.
}\label{ao_astro_AB}
\end{figure*}


From the existence of eclipses with the 94.2-day period we know that the
inclination of this orbit is close to $90^\circ$ ($\sin(i)\simeq1$), so the true $a_1$ 
is close to 0.2~AU. At the distance to the system, this corresponds 
to the projected angular value of the major semi-axis $\hat{a}_1\simeq1.55$~mas. 
This means that we can expect a peak-to-peak astrometric displacement of $\sim$3~mas 
of star A relatively to B, which should be measurable \citep[][and the errors in 
Table~\ref{tab_ao_obslog}]{neu07,rol08,hel09b}. Comparison of two images, taken in 
2013 and 2016, presented on the Figure~\ref{ao_fits}, shows that we see a real 
relative motion between the two stars. The difference of $\sim$20~mas ($\sim$2.6~AU) can not 
be explained by the improper distortion correction (this would increase individual errors), 
nor the atmospheric refraction (measurements from observations in different filters 
from the same night are in agreement, the scale of refraction is much smaller).

In Figure~\ref{ao_astro_AB} we show the measurements of the position of A relatively 
to B on the $\Delta\alpha/\Delta\delta$ plane. The left panel shows all data available
from the WDS (starting from year 1831). Unfortunately the uncertainties are not given or can 
not be estimated in most cases. Nevertheless, one can easily note that in 185 years the two stars 
approached each other, moving with the average speed of 5.26(56)$\times$10$^{-3}$ mas/d or 1.9(2) mas/yr 
in $\alpha$, and 9.55(95)$\times$10$^{-3}$ mas/d or 3.5(3) mas/yr in $\delta$. This seems to be 
the orbital motion, as the measured proper motion is over 2 times larger. One can deduce that in 
$\sim$200~yr the two stars will pass very close to each other. The orbital period is most likely 
of the order of single thousands of years. Assuming $P_{AB}\sim1000$~yr, and taking the estimates 
of total masses of A and B ($m_{AB,tot}\simeq4.32$~M$_\odot$; see next Sections), we can 
estimate the physical major semi-axis to be $\sim$165~AU, or $\sim$1.28~asec in angular units, 
at the distance to the system ($129.4\pm7.7$~pc). This suggests a high-inclination orbit 
and/or significant eccentricity. 

The middle panel of Fig.~\ref{ao_astro_AB} depicts only our Keck~II/NIRC2 astrometry of star 
A relatively to B. The orbital displacement, predicted for the time span of observations, is
plotted over. One can see that the measurements do not lay on the orbital motion path, and
their spread is much larger than 3~mas (predicted for the 94.2-day orbit). The orbital motion
seems to be dominant here, therefore, in the right panel of the same Figure, we show our 
measurements corrected for the orbital motion, and shifted to have the first point in (0,0). 
The spread of the data is still larger than 3~mas, and the points still do not lay along 
a single line, which would be expected for $i\simeq90^\circ$. Moreover, when orbital 
phases are calculated (from the ephemeris given in Sect.~\ref{sec_kep_phot}), it turns out that 
all Keck observations were done in phases 0.1$-$0.3 (as labelled). If only the 94.2-day orbit was 
responsible for the observed positions, one would expect all the measurements to be well within 
$\sim$1.5~mas. A possible explanation is another body orbiting one of the components, but it 
is impossible to say which one, from the relative astrometry alone. It is difficult to estimate 
the parameters of the putative orbit, because the orientation of the one with $P=94.2$~d is unknown, 
and we only have precise measurements from three epochs. More high-precision astrometric data,
from AO or interferometric, are needed to properly model 
all possible orbits. One should also keep in mind, that the long-term, gradual motion of A 
relatively to B (or vice versa) that we found from archival data, is quite uncertain. The archival 
astrometry from the WDS is not precise enough, and in most cases the uncertainties are not available. 
Nevertheless, it shows that the relative position of B vs. A has not changed much for nearly 
two centuries, so our results can not be explained by different proper motion and/or parallax 
of the two components. We also do not see any significant variations in the RV residuals.

\subsection{Absolute parameters of KIC~4150611~B}

\begin{table}
\centering
\caption{Absolute parameters of the components of KIC~4150611~B.}\label{tab_par}
\begin{tabular}{lcc}
\hline \hline
Parameter & Value & $\pm$ \\
\hline
$M_{Ba}$ (M$_\odot$)	& 0.894	& 0.010	\\
$M_{Bb}$ (M$_\odot$)	& 0.888	& 0.010	\\
$R_{Ba}$ (R$_\odot$)	& 0.802	& 0.044	\\
$R_{Bb}$ (R$_\odot$)	& 0.856	& 0.038	\\
$a$ (R$_\odot$)	&21.508	& 0.075	\\
$\log(g_{Ba})$		& 4.581	& 0.048	\\
$\log(g_{Bb})$		& 4.522	& 0.038	\\
$v_{syn,Ba}$ (km/s)$^a$	& 4.69	& 0.26	\\
$v_{syn,Bb}$ (km/s)$^a$	& 5.00	& 0.20	\\
$\tau_{syn}$ (Myr)$^b$	& 56.51	& 0.24	\\
$\tau_{cir}$ (Gyr)$^b$	& 31.21	& 0.09	\\
\hline
\end{tabular}
\\$^a$ Rotation velocities, under the assumption of pseudo-synchronisation.
$^b$ Time scales of spin-orbit synchronisation, and circulation of the orbit.
\end{table}

\begin{figure*}
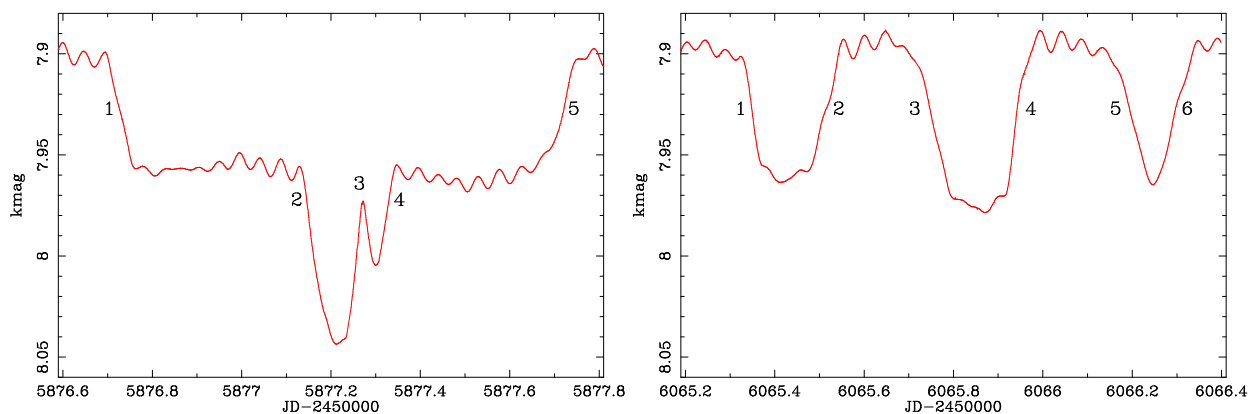

\centering
\includegraphics[width=0.9\columnwidth]{ecl_94_deep.eps}
\includegraphics[width=0.9\columnwidth]{ecl_94_triple.eps}
\caption{Examples of two kinds of eclipses of the F1-type pulsator.
{\it Left:} A ``deep'' eclipse. One star transits the pulsator between
moments 1 and 5, while the other only from 2 to 4. They eclipse each
other in 3.
{\it Right:} A triple eclipse. One star transits the pulsator 
between moments 1 and 2, and later between 5 and 6. The other 
star transits between 3 and 4.
The wave-like modulation comes from pulsations. Note the same 
vertical scale on both panels. The exact depths vary slightly from
one event to the other. Short-cadence data are shown.
}\label{fig_ecl_94}
\end{figure*}

Absolute values of stellar parameters of the components of the 8.65-day pair were 
calculated with the JKTABSDIM procedure, available together with the JKTEBOP. 
This simple code combines several spectroscopic and LC parameters 
(i.e.:~$P,K_{1,2},i,e,r_{1,2}$) to derive a set of stellar absolute dimensions 
($M_{1,2},R_{1,2},a$), and related quantities ($v_{syn1,2},\log(g_{1,2})$).
Using formalism of the theory of tidal interactions, it also predicts the time
scales of spin-orbit synchronisation ($\tau_{syn}$), and circularisation of
the orbit ($\tau_{cir}$). If desired, the JKTABSDIM also calculates radiative 
properties ($L_{1,2}/L_\odot,M_{bol1,2}$) and distance, but requires both effective 
temperatures, multi-colour photometry, and $E(B-V)$ as input. Due to lack of such 
data, we did not attempt to estimate the distance, but instead rely on the {\it Gaia} 
parallax \citep{gai16}. 

The parameters are presented in Table \ref{tab_par}. Indices `$Ba$' and `$Bb$' are
used instead of `1' and `2'. We reached 
$\sim$1.1 per cent precision in masses, and 4.6-5.5 per cent precision 
in radii. The latter is hampered by additional photometric variability in the system, 
presence of the (dominant) third light, and the fact that in JKTEBOP analysis we did 
not use spectroscopic flux ratios, which help to constrain the ratio of the radii. 
From the parameters from Tables \ref{tab_par_sb2}, \ref{tab_0415_kk}, and \ref{tab_par},
one can see that the components of the studied pair are nearly identical,
both being smaller, and slightly less massive than the Sun, and the more
massive component (here, the primary Ba) seems to be smaller. All the important
ratios -- mass, radii, and fractional fluxes -- agree with unity within errors.
The contribution to the total system's flux from this binary is 14.12(45) per cent,
with 6.86(6) and 7.25(6) per cent individual contributions from the primary and secondary,
respectively.

\subsection{The 94.2 and 1.52-day periods.}

\begin{figure}
\centering
\includegraphics[width=0.95\columnwidth]{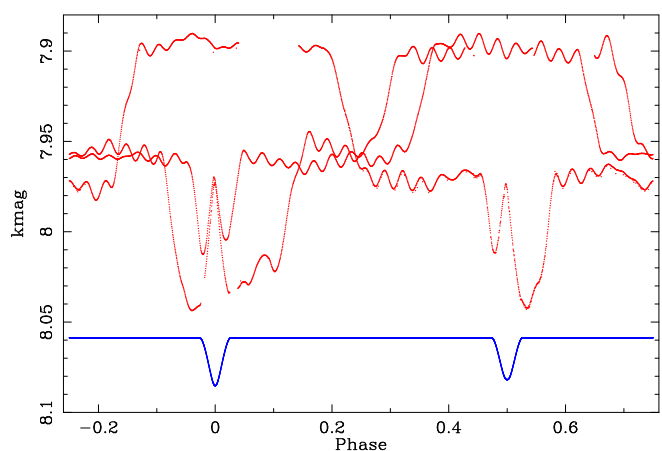}
\caption{Three ``deep'' eclipses of the F1-type pulsator (red points,
short-cadence data) phase folded with the period of the
1.52-day pair. The model curve of this pair is shown in
blue (scaled and shifted for clarity). The brightening events in the ``deep''
eclipses coincide with the eclipses of the 1.52-day par, 
proving that this is the system revolving around the pulsator.}\label{fig_94_vs_15}
\end{figure}

In Section~\ref{sec_hyb} we discussed the orbital motion of the component A
(F1-type pulsator) with the 94.2-day period. As it was mentioned before,
the related eclipses have quite peculiar shapes, which is due to the binary 
character of the object revolving around this star. Those eclipses come in two
shapes: ``triple'' with three small dips, or ``deep'' with one long drop in brightness,
which deepens in its central part. Examples of a ``triple'' and a ``deep''
eclipse are shown in Figure~\ref{fig_ecl_94}. 

The ``triple'' eclipse occurs when one star passes quickly in front of the
pulsator, then is followed by a transit of the other star, and again by a 
passage of the first one, which this time is moving in the opposite direction.
In the ``deep'' event, the first star starts to transit the pulsator, but 
the change of the observed direction (due to its orbital motion) occurs when 
it is still in front of the pulsator, therefore this transit is relatively
long. During this eclipse the second star also transits the pulsator, which
is the cause of the deeper part of the event, when larger area of the eclipsed
component is obscured. Note also that during this deeper part there is a 
small brightening. It happens when the two transiting stars eclipse each other,
and the obscured area of the star behind them is smaller. It is therefore 
easy to associate these brightening events with eclipses that occur with one 
of the other periods. They coincide with the 1.52-day period (see curve S4), 
meaning that this is the eclipsing pair that revolves around the F1-type star on 
the 94.2-day orbit. Figure~\ref{fig_94_vs_15} shows three ``deep'' eclipses 
(short-cadence data) phase-folded with the ephemeris for the S4 curve from 
Table~\ref{tab_0415_kk}. The brightenings occur exactly at phases 0.0 and 0.5.

Since we have recorded the orbital motion of the pulsator, we also attempted
to do it for the 1.52-day pair, using the ETVs. This is however very difficult, 
as the eclipses are shallower than the $rms$ of the curve. The expected signal
has the amplitude $A_{Ab}$ of the order of single minutes, but the errors of measurements
themselves and their spread have larger values. Therefore we could not detect the 
signal securely, even though the period is known. We can only determine the 
detection limit of $A_{Ab}<283$~s ($rms$ of the ETVs), which translates into the 
limit of RV amplitude of the centre of mass $K_{Ab}<66$~km/s 
\citep[from Eq.~8 in][]{hel16}. Together with the RV amplitude of the pulsator
itself ($K_{Aa}=22.6\pm2.2$~km/s), we can determine the limit for ratio of masses:
$q_{A} \equiv (M_{Ab1}+M_{Ab2})/M_{Aa} = K_{Aa}/K_{Ab} \ga 0.34$.

\subsection{The 1.43-day period and the star C}\label{sec_per_1.43}

\begin{figure}
\centering
\includegraphics[width=0.99\columnwidth]{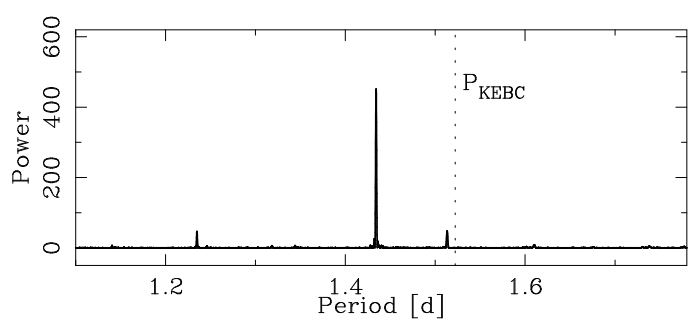}
\caption{A piece of the LS periodogram of the residuals
of the JKTEBOP fit to the S1 curve around the value of 1.5222786~d,
as given in the KEBC (marked with dashed vertical line). Instead of
this value, we found a peak at 1.43420486 d. The LC phase-folded
with this period (and cleared from pulsations) is the S3 curve 
(middle panel of Fig.~\ref{fig_mod_0415}).}\label{fig_per_1.43}
\end{figure}

The last period, not discussed so far, is the 1.4342~d, which is the period
of the most prominent brightness variation in the S3 curve (Fig.~\ref{fig_mod_0415}).
Because it was not listed in the KEBC, we were not aware of its existence, 
until we run the periodogram on the residuals of the JKTEBOP fit to the 
S1 curve. We were looking for the 1.52-day period, instead we noted the 
peak at $\sim$1.43 (Figure~\ref{fig_per_1.43}).

It is difficult to associate it with any of the two bright visual 
components. The \kep pixel mask of the target is elongated and relatively large 
(due to the brightness), therefore covers several nearby faint stars, but 
changes from quarter to quarter and the common area is small. Because the
eclipses are seen in all quarters, the source must be close to the bright
visual pair. Therefore we suspect that the third star seen in the AO
images is also an eclipsing binary, and 1.4342~d is its orbital period.

\begin{figure}
\centering
\includegraphics[width=0.99\columnwidth]{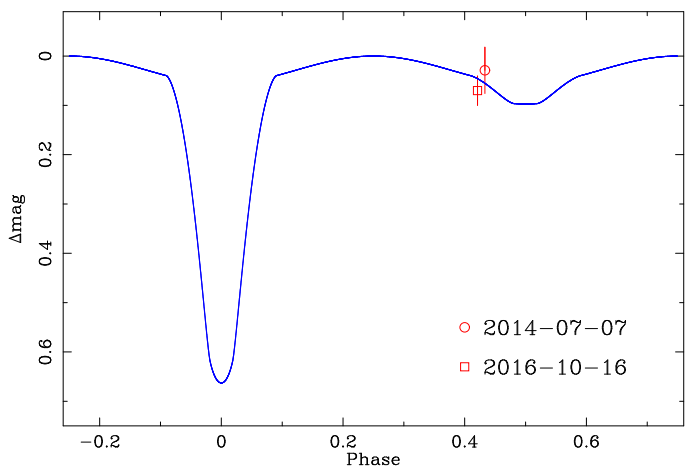}
\caption{Photometric measurements of the star C, relatively to star A in Kcont
(red symbols) as a function of the orbital phase of the S3 curve. Plotted over
is the model S3 curve, corrected for the third light. The two measurements
are shifted by 5.11 mag to match the model curve at the phases of observations.
The unfortunate timing of the runs did not allow us to
verify if the star C is the S3 eclipsing binary, even if the expected 
secondary eclipse is deeper in Kcont than in the \kep band.}\label{fig_ao_photo}
\end{figure}

In order to confirm that, we performed photometric measurements on the
archival and our own AO observations. We found out that the data from 2014
July 7 were taken during the beginning of the secondary eclipse, which helped
us choose the observing set-up for our own observations. Before our run,
we performed preliminary photometric measurements, and found that the lowest
$rms$ is reached in Hcont and Kcont filters. We decided to observe in Kcont,
using exactly the same settings as in 2014, because the expected brightness
variation was larger than for Hcont (for eclipsing binaries composed
of stars of significantly different $T_{eff}$, the longer the wavelength the
deeper is the secondary eclipse). Additionally, in October we decided to 
observe the target also in the K (clear) filter. This set-up is more convenient
and time efficient that the previous one, which includes a neutral density 
filter, and requires longer integrations and calibrations.

Unfortunately, the night of 2016-09-11, when
the model predicted maximum brightness of the S3 curve, the conditions were too 
poor, and quality photometric measurements of the star C were impossible.
To make matters worse, the orbital phase during the other night (2016-10-15) 
was almost exactly the same as in 2014, so no significant brightness 
variation is observed (Fig. \ref{fig_ao_photo}). Therefore, with our current 
data, we can not confirm that the star C is the eclipsing pair with the period
of 1.4342 d, but we can not exclude it either. Additional observations
are required, and they should optimally be taken with the K+clear set-up.

\begin{figure}
\centering
\includegraphics[width=0.99\columnwidth]{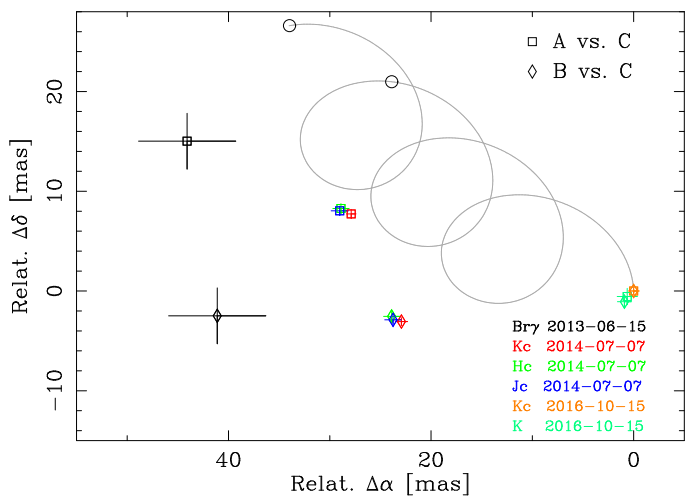}
\caption{Astrometric measurements of stars A (squares) and B (diamonds)
relatively to C. Measurements are shifted so that the one in Kcont from 2016 October
is on (0,0). Each measurement (date, filter) is denoted with a different
colour. The grey line shows the apparent motion on the sky over the course
of the observations, expected from the parallax and proper motion values
from {\it Gaia} DR1, and if the star C was a steady background object.
Black circles mark the moments of archival observations (2013 June and
2014 July). These measurements show that the star C is probably not
bound gravitationally with the AB pair, but may have a measurable proper
motion. The difference in observed paths of stars A and B comes from their
relative measurements, discussed in Sect.~\ref{sec_hyb}.
}\label{fig_ao_astro}
\end{figure}

We also attempt to verify if the star C is bound to the AB system. In 
Figure~\ref{fig_ao_astro} we show how the measured positions of stars A and B
relatively to C change over time (grey line), assuming that C is a distant background object
that has no measurable proper motion. Our measurements do not lay on the path 
predicted by the parallax and proper motion of KIC~4150611 from {\it Gaia}~DR1,
but they also do not agree with C being gravitationally bound to AB. We already
discussed the possible systematics in Sect.~\ref{sec_hyb}. We conclude that C
is most likely a background object, but has a measurable proper motion of few mas/year.

\begin{figure}
\centering
\includegraphics[width=0.99\columnwidth]{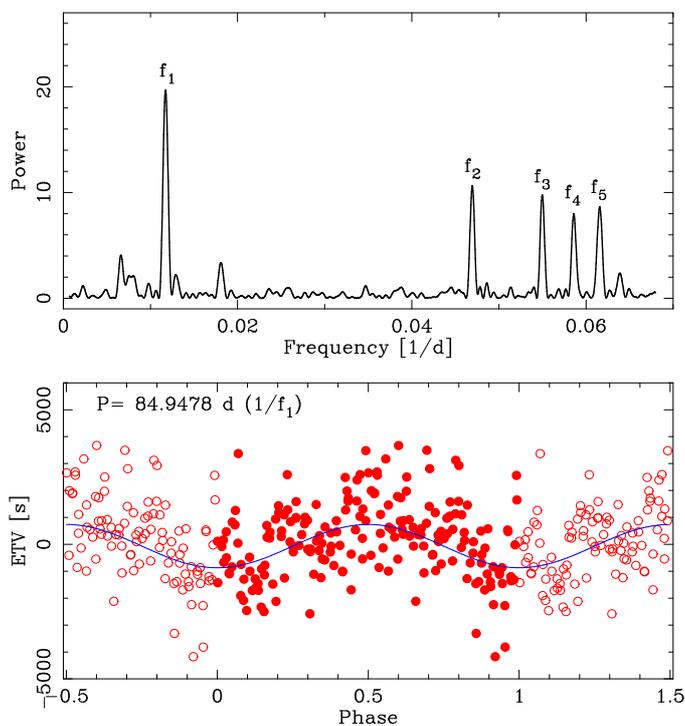}
\caption{{\it Top:} An LS periodogram (in frequency domain) of the ETVs
of the S3 curve. Five most prominent frequencies are marked.
They were also detected by FREDEC.
{\it Bottom:} The ETVs (red points) phase-folded with the period corresponding
to the highest peak of the periodogram. The best-fitting sine function is plotted over 
(blue line). The data are repeated at phases $<0.0$ and $>1.0$ for clarity.
}\label{fig_etv_EB2}
\end{figure}

\begin{table}
\centering
\caption{Frequencies identified in the LS periodogram and by FREDEC in the 
ETVs of the S3 curve, including their false alarm probability (FAP) and ratios 
(last column). High-order ratios of $f_3:f_1$ and $f_5:f_1$ can be accidental 
and should be treated with caution.}\label{etv_fred}
\begin{tabular}{ccccr}
\hline \hline
$n$	& $f_n$	& $P_n = 1/f_n$	& FAP	& $f_n : f_1$ \\
	& (1/d)	&	(d)	&	&	\\
\hline
1	& 0.011771 & 84.9478 & 1.79$\times10^{-8}$ & 1:1 \\
2	& 0.046988 & 21.2818 & 2.02$\times10^{-6}$ &$\sim$4:1 \\
3	& 0.054928 & 18.2057 & 7.06$\times10^{-5}$ &$\sim$14:3 \\
4	& 0.058641 & 17.0523 & 1.06$\times10^{-4}$ &$\sim$5:1 \\
5	& 0.061626 & 16.2271 & 3.23$\times10^{-4}$ &$\sim$21:4\\
\hline
\end{tabular}
\\FREDEC FAP of the whole quintuple: 1.06$\times10^{-4}$.
\end{table}

Finally, we have checked the ETVs of S3, obtained as described in Section~\ref{sec_etv}.
After removing 3 outliers, we run an LS periodogram on our 197 measurements.
Their mean separation is $\sim$7.35~d, therefore we looked for periods longer than
14.7 days. We detected 5 candidate peaks, with the strongest one at $P\sim85$~d. 
We immediately noticed that frequencies of at least two others 
are multiples of the strongest one. To confirm the significance of the set of 
frequencies we found, we run a multi-frequency periodogram with the FREquency 
DEComposer algorithm \citep[FREDEC;][]{bal13a,bal13b}. We found that this combination 
of peaks is statistically significant, with all having false alarm probability 
below 0.05\%. These results are presented in Table~\ref{etv_fred} and 
Figure~\ref{fig_etv_EB2}.

In Fig.~\ref{fig_etv_EB2} we also show our ETVs phase-folded with the 85-day
period, and the best-fitting sine function. Despite the large scatter, the modulation 
with the amplitude of 804$\pm$85~s is clearly seen, even if it is lower than the $rms$
of the fit (1281~s). The amplitudes of other periods are no larger than 500~s, and when
fitted for, the $rms$ drops only to 1200~s. It is difficult to assess if these four 
frequencies are physically real, or just some sort of artefacts in the data. 
With current sampling, each subset of $\sim$7.35~d covers about 5 orbital periods
of the EB. Given the observed depth (or shallowness) of the eclipses, this seems
to be a feasibility limit of the method we used -- individual ETV measurement errors are
quite large already. With longer sampling (e.g. 100 subsets, 14.7~d mean cadence), the
ETV errors are obviously smaller. The 85~d period is still very well visible, but 
the other four become shorter than the approximate Nyquist cut-off ($\sim$29.4~d) and
their secure detection is impossible. We then treat only the longest period as realistic.

However, assuming so, and that the modulation is caused by another body orbiting the EB, 
the observed amplitude can be translated into an unrealistic value of the mass function
$f(M) \simeq 77.5$~M$_\odot$. Such a configuration would be difficult to explain. 
Therefore, the ETVs we observe for the S3 curve are rather not caused by another body.
They might possibly reflect a different phenomenon, occurring on one of the
stars of the whole KIC~4150611 system, like evolution of spots, which we know exist on the
Ba+Bb pair, because we see the rotational period in the periodogram in Fig.~\ref{fig_per0415}.
They may also be a result of a systematic factor we did not take into account.

\subsection{Comparison with isochrones}

\begin{figure}
\centering
\includegraphics[width=0.9\columnwidth]{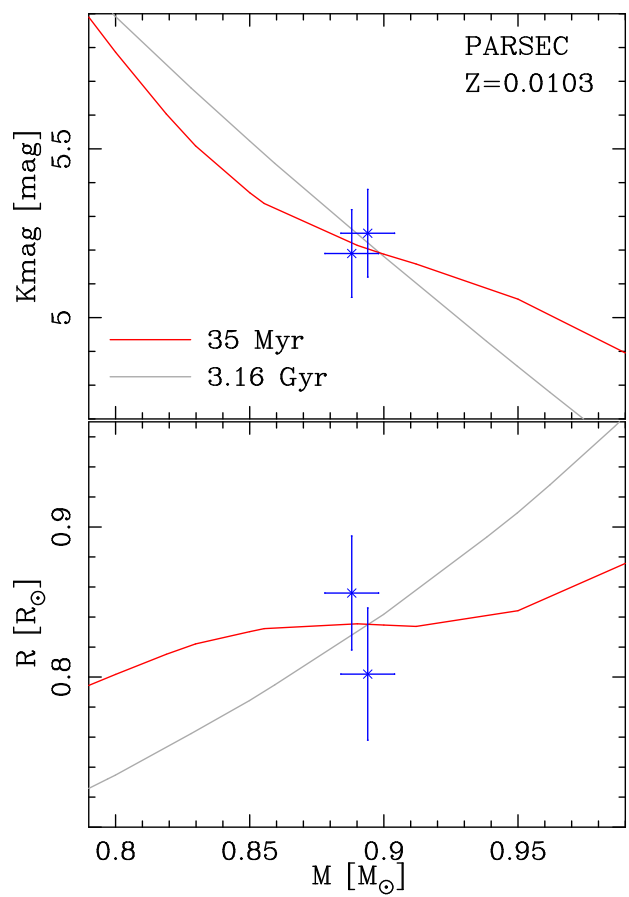}
\caption{Comparison of our results for the G-type eclipsing
pair with theoretical PARSEC $Z=0.0103$ isochrones on mass vs. 
absolute \kep magnitude (top) and radius (bottom) planes. A good
agreement is found for ages of 35~Myr (red line) and 3.16~Gyr 
(grey line). Stellar parameters predicted for the F1-type pulsator
favour the former age.}\label{fig_iso}
\end{figure}

We compare our results, i.e. masses, radii and absolute magnitude, with the 
theoretical PARSEC isochrones \citep{bre12}, that include calculation of 
absolute magnitudes in the \kep photometric band. The estimate of iron 
abundance from \citet{nie15} is $\log \epsilon(\mathrm{Fe})=7.33\pm0.10$.
Assuming the customary logarithmic abundances scale, with $\log \epsilon(\mathrm{H})=12$,
and solar iron abundance of $\log \epsilon(\mathrm{Fe})_\odot=7.50\pm0.04$
\citep{asp09}, we get the value of $[Fe/H]=-0.17\pm-0.11$. We assume that
this represents also the metallicity $[M/H]$ of KIC~4150611, and use PARSEC
isochrones for this $[M/H]$ value, which translates into $Z\simeq0.0103$
for this set.

Comparison of our results for Ba+Bb with the models on the $M/R$ and $M/Kmag$ 
planes shows a good agreement with the 35~Myr isochrone (Figure~\ref{fig_iso}). 
For the estimated absolute magnitude of the pulsating star Aa ($2.52\pm0.13$~mag), 
we can find its theoretical mass, radius, $\log(g)$, and temperature 
-- 1.64(6)~M$_\odot$, 1.376(13)~R$_\odot$, 4.38(1)~dex, and 8440(280)~K, 
respectively. This is in disagreement with \citet{nie15}, who give 3.8(2)~dex
and 7400(100)~K, but it may be at least partially explained by uncertainties 
in the metallicity and age determinations, or the influence of additional flux on 
the spectral analysis. This may also be a hint for the existence of additional, 
relatively bright source in the system, i.e. that the pulsator constitutes less 
than $\sim$85 per cent of the total flux. In such situation the isochrone-predicted 
$T_{eff}$ would be lower.

Another relatively good fit is found for a 3.16~Gyr isochrone, but in such case the 
component Aa would be an F5-F6 type star ($5000<T_{eff}<6600$~K), and this is in strong
disagreement with any spectral type estimation for this star. It would also have a 
mass of 1.32--1.36~M$_\odot$. We find this scenario to be unlikely for a $\delta$~Sct 
pulsator, and adopt 35~Myr as the age of the system.

Taking our isochrone-based estimation of the mass of the star Aa, and the mass function 
value derived from the RVs (0.113$\pm$0.033~M$_\odot$), we can estimate that the total
mass of the Ab1+Ab2 pair is 0.90(13)~M$_\odot$ (assuming $\sin^3(i)\simeq1$).
It is in a reasonable agreement with masses predicted by the isochrone from our
$Kmag$ estimates, which are 0.44$\pm$0.13 and 0.32$^{+0.21}_{-0.06}$~M$_\odot$ for the
primary and secondary, respectively (total of 0.77$^{+0.25}_{-0.14}$~M$_\odot$). 
The agreement would be slightly better if the component Aa was fainter (cooler and less 
massive), which may be another hint for an additional flux in the \kep LC.

At the assumed age of 35~Myr, the PARSEC isochrone predicts that the Ab1+Ab2 pair 
would be still in pre-main-sequence stage, with radii of 0.61$\pm$0.09 and 
0.52$^{+0.14}_{-0.05}$~R$_\odot$. It is therefore plausible that this pair is 
still very active, and responsible for the relatively weak X-ray emission detected by
ROSAT \citep{vog00}. Taking the isochrone-predicted masses and the orbital
period 1.5222468(25)~d, we can estimate the major semi-axis of the orbit:
$a_{Ab}=5.11^{+0.55}_{-0.31}$~R$_\odot$. This leads to fractional radii of 
$r_{Ab1}=0.119_{-0.021}^{+0.019}$ and $r_{Ab2}=0.102^{+0.033}_{-0.017}$. They are 
in agreement with the results of the JKTEBOP fit to the S4 curve (Table~\ref{tab_0415_kk}).
The ETV signal of this pair, produced by the orbital motion around the component Aa 
with $P=94.226$~d, would be $\sim$180-210~s, which is below our detection limit. 

From the isochrones, we can also estimate effective temperatures of Ba and Bb, 
to be 5680 and 5640~K, respectively. This is consistent with the observed G spectral 
type. The age of 35~Myr suggests that this pair has not reached the state
of pseudo-synchronisation yet. The theoretical time scale for its masses and
period is $\tau_{syn}\simeq57$~Myr (as calculated by JKTABSDIM). Therefore, 
the peaks in the periodogram at frequencies 0.18-0.19~d$^{-1}$ (Fig.~\ref{fig_per0415}) 
can be explained by a super-synchronous rotation of one or both components. 
In a pseudo-synchronous case, the frequency would be $\sim0.11$~d$^{-1}$
($P_{rot}\simeq8.7$~d).

\subsection{Galactic kinematics}
To verify the young age of the system, we checked its Galactic velocity. 
Using the known parallax and proper motion (Tab.~\ref{tab_prop}) and our 
value of the systemic velocity $v_\gamma$ for the component B (the more 
reliable one, we have calculated the spatial motion components: $U=2.52\pm0.28$, 
$V=-23.29\pm0.16$ and $W=-2.49\pm0.13$~km/s (no correction for the solar movement  
has been done). These values put KIC~4150611 well within the thin disk, probably in 
a moving group called Coma Berenices or ``local'' \citep{nor04,fam05,sea07}.
\citet{fam05} have shown that ages of stars from this group vary from several
to few hundreds of Myr. This confirms our ``young'' isochrone age of 35~Myr, and
makes the ``old'' one (3.16~Gyr) even less probable.


\section{Summary and future prospects}

\begin{figure}
\centering
\includegraphics[width=\columnwidth]{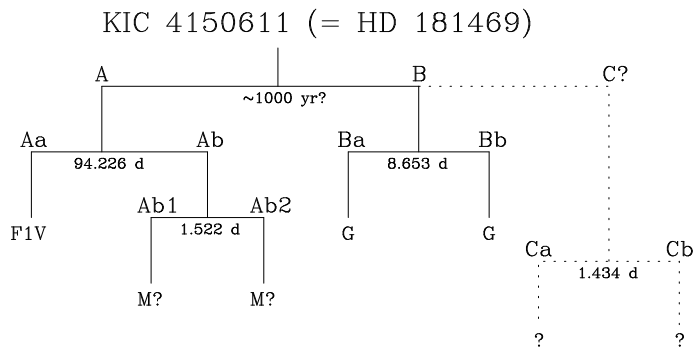}
\caption{A mobile diagram showing the structure of the KIC~4150611 multiple
system. The secure quintuple configuration is drawn with solid lines, orbital periods
and spectral types (or their estimates) are shown. Our uncertainty of the character
and membership of source C to the system is represented by dotted lines and question 
marks on its branch. The putative additional body on an orbit around either star
A or B, discussed in Sect.~\ref{sec_ao_AB}, is not shown.}\label{fig_mob}
\end{figure}

In Figure \ref{fig_mob} we present the configuration of the multiple system
KIC~4150611. The orbital periods and spectral types (at the ``ends'' of each branch)
are given to the best of our knowledge. The uncertain character of the source C is
taken into account. The additional body that could explain the AB pair astrometric 
measurements (Sect.~\ref{sec_ao_AB}) is not shown.

By analysing the eclipses seen in the \kep light curve, HIDES radial velocities
and AO observations from the Keck~II telescope, we obtained a consistent image of a
bright, interesting multiple system KIC~4150611. We managed to directly measure
physical parameters of two of the components (Ba and Bb), which allowed us to find that
this is a relatively young system. From its age we inferred properties of three 
other component (Aa, Ab1 and Ab2). Our results are still incomplete though.

Detailed analysis of the currently available data still leaves some open questions:
\begin{enumerate}
\item What is the 1.43-d eclipsing binary?

The faint star C seems to be a good candidate, but we can not confirm it. We lack 
photometry taken during the primary (deep) eclipse. Sufficient observations do not 
have to be made with AO facilities, but good seeing conditions and a relatively large
mirror is necessary to resolve the star from the AB pair, and obtain sufficient 
SNR of the target. If the star C is not the 1.43-d eclipsing binary, then another 
source must exist in the vicinity of the AB pair.

\item What is the proper motion of the system?

The star C seems to be a distant background object, so a good reference point for 
astrometric measurements of the AB pair, but our measurements do not match the
{\it Gaia} DR1 results. The faintness of C suggests a large distance, so a very
slow proper motion would be expected. The discrepancy may be caused by an 
incorrect determination of $\mu_\alpha$ and $\mu_\delta$. Hopefully, this will
be clarified with future data releases from {\it Gaia}.

\item Where does the 85-day period in ETVs come from?

The signal in our ETVs of the 1.43-d binary is clear and statistically significant,
but it leads to an improbable physical configuration (very high mass function). 
It is thus not clear if this modulation is an artefact, or has a physical origin, like
evolution of spots. Confirmation would come from further timing measurements, but the
source has to be resolved from the bright components A and B. This will only be possible
if the star C is the 1.43-d binary. Other periods we found in the ETVs also require 
confirmation.

\item What is the true number of components?

We presented several hints suggesting that another body may exist in the system. 
First, the isochrone-predicted properties (like temperature of Aa or masses of 
Ab1+Ab2) would be more consistent with observational constraints if the 
pulsator was fainter than what we found from our JKTEBOP fits. Second, as mentioned 
above, the 1.43-day period may not originate from the (seemingly unrelated) star C.
Finally, our relative astrometry of A and B show a motion that can not be explained 
with the 94.2-day orbit. To verify this, more quality AO observations are required, 
including coronagraphic images of the surroundings of stars A and B. Precise 
($\sim$1~mas level) astrometry from speckle observations, and optical or infra-red 
interferometers, is also welcome. Confirmation of the existence of another body 
would make KIC~4150611 a sextuple, or, still possibly, a septuple, which would be 
only the third known case.
\end{enumerate}

Intriguing is the fact that in the system we see four periods of eclipses, with at 
least three coming from the system itself. The triple sub-system A may have two co-planar 
or nearly co-planar orbits (94.2 and 1.52 days), as it is observed in other objects with 
similar architecture, e.g. KOI-126 \citep{car11}, HD~181068 \citep{der11,bor12}, or 
KIC~2856960 \citep{lee13}. The pairs with other two periodicities (8.65 and 1.43 days) 
do not, however, need to share the same orientation. Their inclinations are only calculated 
relatively to the plane of the sky, so the true orientations of their orbital angular momenta 
can still be very different. Astrometric detection of the 94.2-d motion, and complete modelling 
of the orbits and eclipses in the component A, will give valuable information about the 
distribution of momenta in this sub-system, giving important insight into the history of 
formation of this (young!) object.

KIC~4150611 is one of the most interesting astrophysical discoveries of the \kep 
mission, and we believe it deserves further attention. Additional insight into
the evolutionary status and structure of the Aa component may come from detailed
asteroseismic studies of its pulsations. Please note that \citet{shi12} and
\citet{bal14} focus only on its orbital motion, and use only the $\delta$~Sct
pulsations, while \citet{uyt11} only give the classification as a  hybrid. A proper
asteroseismic study is still missing.
 
We would also like to encourage the community to perform new AO observations, in
order to answer the questions described above.


\begin{acknowledgements}
We would like to thank Prof. Andrzej Pigulski from the Astronomical Institute of the 
Wroc{\l}aw University, and Prof. Krzysztof Go\'{z}dziewski from the Toru\'n Centre for 
Astronomy of the Nicolaus Copernicus University for fruitful discussions and valuable 
suggestions, and Dr. Andrei Tokovinin from the Cerro Tololo Inter-American Observatory
for valuable comments and corrections.

This research has made use of the Keck Observatory Archive (KOA), which is 
operated by the W. M. Keck Observatory and the NASA Exoplanet Science Institute (NExScI), 
under contract with the National Aeronautics and Space Administration. Some of the data 
presented herein were obtained at the W.M. Keck Observatory, which is operated as a 
scientific partnership among the California Institute of Technology, the University of 
California and the National Aeronautics and Space Administration. The Observatory was 
made possible by the generous financial support of the W.M. Keck Foundation. 
This work has made use of data from the European Space Agency (ESA)
mission {\it Gaia} (\url{http://www.cosmos.esa.int/gaia}), processed by
the {\it Gaia} Data Processing and Analysis Consortium (DPAC,
\url{http://www.cosmos.esa.int/web/gaia/dpac/consortium}). Funding
for the DPAC has been provided by national institutions, in particular
the institutions participating in the {\it Gaia} Multilateral Agreement.
This research has 
made use of the SIMBAD database, operated at CDS, Strasbourg, France. The authors 
recognize and acknowledge the very significant cultural role and reverence that the summit 
of Maunakea has always had within the indigenous Hawaiian community. We are most 
fortunate to have the opportunity to conduct observations from this mountain.

KGH acknowledges support provided by the Polish National Science Center through grant 2016/21/B/ST9/01613,
and by the National Astronomical Observatory of Japan as Subaru Astronomical Research Fellow. 
This work is supported by the Polish National Science Center grant 2011/03/N/ST9/03192, 
by the European Research Council through a Starting Grant, 
by the Foundation for Polish Science through ``Idee dla Polski'' funding scheme, 
and by the Polish Ministry of Science and Higher Education through grant W103/ERC/2011.
CB acknowledges support from the Alfred P. Sloan Foundation.

\end{acknowledgements}

\begin{appendix} 
\section{RV measurements}

In Table~\ref{aptab_rv} we present single measurements of RVs of components Aa, Ba, and Bb.
For Aa, for every observation, we initially assumed equal uncertainties of 6 km/s. For Ba and Bb
they were obtained from our TODCOR runs. Final measurement uncertainties $\epsilon$, given in 
the Table, are scaled to have the reduced $\chi^2$ of the fit close to 1.

\begin{table}
\centering
\caption{Radial velocity measurements used in this work.\label{aptab_rv}}
\begin{tabular}{lrrrrrrrrr}
\hline \hline
HJD & $v_{Aa}$ & $\epsilon_{Aa}$ & $(O-C)_{Aa}$ & $v_{Ba}$ & $\epsilon_{Ba}$ & $(O-C)_{Ba}$ & $v_{Bb}$ & $\epsilon_{Bb}$ & $(O-C)_{Bb}$ \\
-2450000 & (km/s) & (km/s) & (km/s) & (km/s) & (km/s) & (km/s) & (km/s) & (km/s) & (km/s) \\
\hline
6866.212043 & -23.37 & 2.34 &  2.92 & -57.133 & 0.168 &  0.038 &   11.881 & 0.182 &  0.231 \\
6866.979087 & -25.09 & 2.34 &  0.05 & -64.910 & 0.170 & -0.260 &   19.199 & 0.180 &  0.010 \\
6867.251395 & -24.92 & 2.34 & -0.19 & -65.725 & 0.323 & -0.165 &   20.110 & 0.258 &  0.004 \\
6868.065963 & -23.97 & 2.34 & -0.47 & -64.250 & 0.187 & -0.092 &   18.723 & 0.215 &  0.031 \\
6868.184130 & -24.64 & 2.34 & -1.31 & -63.338 & 0.198 &  0.146 &   17.865 & 0.173 & -0.148 \\
6870.002952 & -21.93 & 2.34 & -1.30 & -35.688 & 0.265 &  0.067 &  -10.252 & 0.201 & -0.320 \\
6913.990303 & -27.24 & 2.34 & -0.81 & -10.650 & 0.202 &  0.019 &  -35.018 & 0.155 &  0.192 \\
7062.376385 & -12.46 & 2.34 &  2.67 &  62.588 & 0.215 & -0.260 & -109.603 & 0.201 & -0.359 \\
7111.244355 & -34.25 & 2.34 &  4.59 & -55.916 & 0.207 &  0.159 &   10.253 & 0.272 & -0.294 \\
7148.151220 & -27.23 & 2.34 &  0.17 &  16.566 & 0.220 & -0.073 &  -62.343 & 0.210 &  0.376 \\
7302.007666 & -43.21 & 2.34 & -1.69 & -50.414 & 0.203 & -0.375 &    4.748 & 0.196 &  0.285 \\
7526.176108 & -30.46 & 2.34 & -4.74 & -60.580 & 0.268 &  0.151 &   15.270 & 0.157 &  0.032 \\
7530.264186 & -16.78 & 2.34 &  2.86 &  57.936 & 0.350 &  0.198 & -104.312 & 0.447 & -0.210 \\
7539.110077 &  -9.35 & 2.34 & -0.74 &  43.984 & 0.508 &  0.325 &  -89.638 & 0.345 &  0.294 \\
7669.010773 & -28.63 & 2.34 & -0.38 &  34.977 & 0.230 & -0.060 &  -81.013 & 0.198 &  0.240 \\
7671.067542 & -31.93 & 2.34 & -0.63 & -58.882 & 0.200 &  0.008 &   13.425 & 0.259 &  0.042 \\
7672.956861 & -35.00 & 2.34 & -0.99 & -63.287 & 0.369 & -0.025 &   17.751 & 0.361 & -0.038 \\
\hline
\end{tabular}
\end{table}

\end{appendix}

\end{document}